\documentclass[12pt]{iopart}
\def\as{\alpha_s}
\def\a0{\bar\alpha_0}

\def\aPT{\as^{\mbox{\tiny PT}}}
\def\ee{e^+e^-}
\def\eps{\epsilon}
\def\cA{{\cal A}}

\def\mI{\mu_{\mbox{\tiny I}}}
\def\mR{\mu_{\mbox{\tiny R}}}
\def\MSbar{{$\overline{\mbox{\rm MS}}$}}

\def\re#1{(\ref{#1})}
\def\beq{\begin{equation}}   \def\eeq{\end{equation}}
\def\beeq{\begin{eqnarray}}   \def\eeeq{\end{eqnarray}}
\usepackage{iopams} 
\usepackage{epsfig} 
\begin{document}

\title{Power Corrections in QCD}

\author{Mrinal Dasgupta\dag\
\footnote[3]{dasgupta@mail.desy.de}
}

\address{\dag\ DESY Theory Group, Notkestrasse 85, Hamburg 22805, Germany.}

\begin{abstract}
This article provides a review of progress made in understanding inverse power law corrections $1/Q^{p}$ where $Q$ is the hard scale involved in a given 
QCD process. Particular attention will be paid to HERA results.

\end{abstract}



\maketitle

\section{Introduction}
In order to test the theory of strong interactions (QCD) it has long been common practice to confront experimental data with perturbative predictions and then extract fundamental parameters such as the strong coupling $\alpha_s$.
A consistent extraction of $\alpha_s$ from different experiments and different observables points at the correctness of assuming QCD as the underlying theory.

Since QCD is a theory where non-perturbative effects play a clear role 
one has to first question the feasibilty of such an approach and review the role of perturbative QCD (pQCD). The fact that one can make perturbative predictions to any extent is indeed a blessing and in the case of many observables 
factorisation theorems allow the separation of pQCD from non-perturbative effects. 
Simply stated a factorisation theorem implies for an observable $V$
\begin{equation}
V(x,Q^2) = (1+C_1 \alpha_s + C_2 \alpha_s^2 +\cdots)\otimes O  +\cdots
\end{equation}
In the above we indicate by $x$ any set of phase space variables on which the observable may depend and by $Q^2$ the hard momentum scale.
$O$ is typically a universal operator matrix element which contains non-perturbative 
information and 
whose value can be extracted from experiment and tabulated, for example a parton distribution measure.
$C_1$ and $C_2$ are perturbative coefficients calculated by using Feynman graphs to represent the hard process.

The dots indicate the limitations of the perturbative approach.
The dots in the brackets tell us that there are missing terms in the perturbative series since any perturbative predictions (even resummed ones) are missing higher order pieces which  
cannot be easily computed. While this is just a technological limitation 
in the sense that higher order calculations become rapidly prohibitive computationally, it is a common misconception that given adequate computing power one could have a well defined perturbative answer that would be arbitrarily accurate. 

That this belief is ill-founded can most easily be seen by examining a particular type of contribution (renormalon graphs) (see for examples and further references \cite{Beneke}) which give that for large enough $n$ the coefficients $C_n$ diverge factorially.
This does not mean that the concept of a perturbative expansion is useless. 
What it implies is that pQCD gives an asymptotic approximation to the observable $V$. The error made in truncating the perturbative expansion is of the order of the first omitted term. Therefore one should carry out the perturbative 
expansion till the smallest term is reached. 
The optimal error which is of the order of this smallest term is thus an 
unvoidable inherent ambiguity in the perturbative expansion 
and one finds 
\begin{equation}
\delta V^{PT} \sim \frac{A}{Q^p}
\end{equation}
where $A$ and $p$ depend on the observable and $\delta V^{PT}$ represents 
the uncertainty of the perturbative estimate
. The coefficient $A$ may contain dependence on one or more phase space variables on which $V$ depends.
 
Also in Eq.~(1) there are dots to indicate that the simple 
factorisation may not be correct at some level, e.g higher twist. These contributions are also typically power behaved and the expectation is that 
in order to obtain an unambigous answer the ambiguity of the perturbative (PT)
 part is cancelled by an ambiguity in the definition of the non-perturbative (NP) higher twist operators.
Assuming further that the size of the ambiguity is proportional to value of NP terms (condensates) one can use the ambiguity in the PT piece to model NP effects which are of a power law type.
In other words where there is a large power behaved ambiguity in the PT series it can be treated as a signal that NP effects are damaging in proportion.
This assumption is known as ultraviolet dominance of higher twist.

Having arrived at the notion that the ambiguity of the perturbative expansion is relevant to the study of power corrections, one is still faced with the task of computing it and comapring with experimental data.
This is conveniently done in the framework of the dispersive approach we shall next discuss.

\section{Dispersive approach}
This method was formulated by Dokshitzer, Marchesini and Webber \cite{DMW} and relies on the assumption  that the power correction to an observable is given 
by evaluating the perturbative renormalon graphs and estimating the ensuing ambiguity of the PT series. 
To investigate this method one observes that inserting renormalon bubbles on gluon lines in a lowest order Feynman graph for a given observable $V$ results in the appearance of a running coupling constant.
If this gluon is an internal line inserting renormalon bubbles 
gives for an observable $V$
\begin{equation}
\label{master}
V^{virt} = \int_0^{\infty} \alpha_s(k^2) \mathcal{V}(k^2) \frac{d k^2}{k^2}
\end{equation}
with $k^2$ representing the gluon virtuality. The role of the renormalon bubbles is simply to give rise to the running coupling in the integrand above.
Strictly within a perturbative context the above integral gives a divergent result due to the divergence of $\alpha_s$ at the Landau pole. 
This is just the factorial renormalon behaviour discussed previously.

To generate power corrections from the above contribution 
the assumption one makes is that the QCD coupling $\alpha_s$ instead of diverging is a well behaved universal 
function down to small scales, i.e infrared finite. Further knowledge of the exact behaviour of the coupling is not required for us to proceed. 
This is certainly not an assumption that can be justified from first principles. However it leads to simple results that can be easily tested against a wealth of experimental data. This exercise has been carried out for different observables and so far it seems that experimental data are in support of this notion of a universal IR finite coupling.
 
Assuming that the coupling is analytic except a 
branch cut across the negative real axis which is a requirement of causality, 
it follows that one can write the
formal dispersion relation
\begin{equation}
\label{alphas}
\alpha_s(k^2) = -\int_0^\infty \frac{dm^2}{m^2+k^2}\;\rho_s(m^2)
\end{equation}
where the spectral function $\rho_s$ represents the discontinuity
across the cut,
\begin{equation}
\label{rhos}
\rho_s(m^2) = \frac{1}{2\pi i}\mbox{Disc}\left\{\alpha_s(-m^2)\right\}
\equiv \frac{1}{2\pi i}\left\{\alpha_s\left(m^2 e^{i\pi}\right)
-\alpha_s\left(m^2 e^{-i\pi}\right)\right\}\;.
\end{equation}

Non-perturbative effects at long distances are expected to give rise
to a modification in the strong coupling at low scales, $\delta\alpha_s$,
which generates a corresponding modification in the
spectral function :
\begin{equation}
\label{deltas}
  \delta\rho_s(m^2)
= \frac{1}{2\pi i}\mbox{Disc}\left\{\delta\alpha_s(-m^2)\right\}\;.
\end{equation}

From Eqs.~(\ref{master}--\ref{deltas}) it follows that the non-perturbative contribution to $V$, trigerred by a virtual gluon with renormalon insertions 
and an NP modified coupling can be written as 
\begin{equation}
\delta V^{virt} = \int_0^\infty \delta{\rho}_s(m^2) \frac{dm^2}{m^2} [\mathcal{V}(m^2)-\mathcal{V}(0)]
\end{equation}
A similar equation holds for the real emission contribution and hence in what follws we shall remove the superscript and talk about 
the sum of real and virtual pieces.

The quantity $\mathcal{V}(m^2)$ represents the $\mathcal{O}(\alpha_s)$ virtual calculation performed with a finite gluon mass $m$. The gluon mass is just the dispersive variable in Eq.~\ref{alphas} and does not imply that one works with a massive gluon field and spoils gauge invariance.

After some further manipulation one gets 
the following non-perturbative contribution to $V$
\begin{equation}
\label{deltaF}
\delta V = \int_0^\infty \frac{dm^2}{m^2}\,\delta\alpha_s(m^2)
\,\mathcal{G}(m^2/Q^2)
\end{equation}
where, setting $m^2/Q^2 = \epsilon$,
\begin{equation}
\label{Gdef}
\mathcal{G}(\epsilon) = -\frac{1}{2\pi i}\mbox{Disc}\left\{\mathcal{V}(-\epsilon)\right\}\;.
\end{equation}
Since $\delta\alpha_s(m^2)$ is limited to low values of
$m^2$, the asymptotic behaviour of $\delta V$ at large
$Q^2$ is controlled by the behaviour of $\mathcal{V}(\epsilon)$ as
$\epsilon \to 0$.  We see from Eq.~\ref{Gdef} that no terms analytic
at $\epsilon=0$ can contribute to $\delta V$.  On the
other hand for a square-root behaviour at small $\epsilon$,
\begin{equation}
\label{deltaF1}
\mathcal{V} \sim a_V \frac{C_F}{2\pi}\sqrt{\epsilon}
\qquad\Longrightarrow\qquad
\delta V = -\frac{a_V}{\pi}\frac{\mathcal{A}_1}{Q}\;,
\end{equation}
where we defined the coupling moment 
\begin{equation}
\label{cAdef}
\cA_1 = \int_0^\infty \frac{dm^2}{m^2} \; m \; \delta\as(m^2)
\end{equation}
We can express $\cA_1$ in terms of the average value of $\as$
in the infrared region, as follows.
We substitute for $\delta\as$ in Eq.~\re{cAdef}
\beq
\delta\as(m^2) \simeq \as(m^2) - \aPT(m^2)\;,
\eeq
where $\aPT$ represents the expression for $\as$
corresponding to the part already included in the
perturbative prediction. If the perturbative calculation is carried out to second
order in the \MSbar\ renormalization scheme, with renormalization
scale $\mR^2$, then we have
\beq
\aPT(m^2) = \as(\mR^2) + [b\ln(\mR^2/m^2)+k]\,\as^2(\mR^2) 
\eeq
where for $N_f$ active flavours ($C_A=3$)
\beq
b = \frac{11 C_A-2N_f}{12\pi}\;,\;\;\;\;
k= \frac{(67-3\pi^2)C_A-10N_f}{36\pi}\;.
\eeq
The constant $k$ comes from a change of scheme from \MSbar\ to
the more physical scheme \cite{CMW} in which $\as$ is preferably
defined at low scales.
Then above some infrared matching scale $\mI$ we assume that
$\as(m^2)$ and $\aPT(m^2)$ approximately coincide, so that
\beeq\label{A1exp}
\cA_1 &\simeq& \frac{C_F}{\pi}
\int_0^{\mI} dm\,\left(\as(m^2)-
\as(\mR^2) - [b\ln(\mR^2/m^2)+k]\,\as^2(\mR^2)\right)\nonumber\\
&=& \frac{C_F}{\pi}\,\mI\,\left(\a0(\mI)-
\as(\mR^2) - [b\ln(\mR^2/\mI^2)+k+2b]\,\as^2(\mR^2)\right)\;,
\eeeq
where
\beq\label{a0def}
\a0(\mI) \equiv \frac{1}{\mI}\int_0^{\mI}\as(m^2)\,dm\;.
\eeq
Studies of event shapes in $\ee$ annihilation 
suggest that $\a0\simeq 0.5$ for $\mI=2$ GeV, which translates into
a value of $\cA_1\simeq 0.2$ GeV for $Q\sim 20-100$ GeV.

\section{DIS event shapes}

Here we shall concentrate on the phenomenology of power corrections to 
event shape variables measured by the H1 and ZEUS collaborations.
This study is carried out in the Breit frame and the current hemisphere $H_c$ 
is used to define the variables, freeing us from contamination by the proton remnant jet.

Some of the observables studied are defined below
 \begin{itemize}
\item Current jet thrust
This is constructed by summing the modulii of hadron components along the photon axis in the Breit frame. Two normalisations are possible, to $Q/2$ on one hand and to the energy in the current hemisphere (provided this exceeds a certain preset minimum).
\begin{equation}
\tau_Q = 1-T_Q = 1- 2\frac{\sum_{a \in H_c}|p_a.\hat{n}|}{Q}
\end{equation}
\begin{equation}
\tau_E = 1-T_E = 1-\frac{\sum_{a \in H_c}|p_a.\hat{n}|}{\sum_{a \in H_c}E_a} 
\end{equation}

\item Current hemisphere jet mass
\beq\label{rhodef}
\rho=\left(\sum_{a\in H_c}p_a\right)^2/Q^2\;.
\eeq

\item $C$-parameter
The $C$-parameter is
\beq\label{Cdef}
C=6\sum_{a,b\in H_c}|\vec p_a||\vec p_b|\sin^2\theta_{ab}/Q^2\;.
\eeq
\end{itemize}

There are still other possible definitions and variables studied in addition to the ones mentioned above (an important one being the jet broadening) 
but for illustrative purposes it is enough to restrict the discussion to the variables mentioned.

\section{Power corrections}
\subsection{Naive results}
The power corrections for different event shape variables are given by 
following the dispersive recepie.
One first computes the one-loop graphs with a finite gluon mass $\eps Q^2$
For all the variables mentioned above the leading non-analytic behaviour at small $\eps$ is a $\sqrt{\eps}$ singularity. 

This leads to $1/Q$ power corrections of the form of Eq.~\ref{deltaF1}

The coefficients $a_v$ differ from variable to variable and 
one gets
\begin{equation}
a_\rho=-2,\:a_\tau =-4, \:a_C = -6 \pi
\end{equation}
The constant $\cA_1$ is expected to be approximately universal (to about $20 \%$)as observed in $e^{+}e^{-}$ data and with similar values to those found there.

One point that needs to be made in the DIS context is that the leading 
power corrections to most event shapes are independent of the Bjorken variable $x$. 
This comes about since the $1/Q$ effects in question are trigered by a soft gluon, the emission of which does not change the momentum fraction carried by the incoming projectile.
An exception to this is the jet broadening variable where an interplay between this emitted soft gluon and $x$ dependent perturbative radiation gives rise to an $x$ dependent power correction. A complete estimate for the broadening power correction relies on a resummed perturbative calculation which shall be published shortly \cite{prep}.

Another point that needed to be addressed was one first raised by Nason and Seymour \cite{NS}. They pointed out that gluon decay can only be mimimcked by a massive gluon provided one can integrate inclusively over the gluon offspring.
In other words the machinary of the dispersive approach should only strictly be applied to inclusive observables which do not register offspring gluons.
Event shapes are not such variables as the offspring gluon momenta affect the event shape. Hence one needs an improved approach where one explicitly treats gluon decay properly.
 \subsection{Two-loop improvement}
The two-loop calculations that take gluon decay into proper account 
were first performed for $\ee$ observables \cite{eeMil}
Following this similar work was performed for DIS observables $\cite{DWMil}$.
The conclusions could at first sight be described as surprising.
Taking two loop effects into account meant a simple rescaling of the old one-loop coefficients by a universal factor $\mathcal{M} = 1.49 $ for $n_f=3$.
Hence the ratios of power corrections to different observables are still given by the ratios of the one-loop coefficents $a_V$.

This surprising result came about as a result of 
\begin{itemize}
\item Universality of soft gluon radiation.
\item Common geometrical properties of event shapes --they are all linear in the emitted transverse momenta.
\end{itemize}
\subsection{Differential distributions}
\begin{figure}
\begin{center}
\epsfig{file = 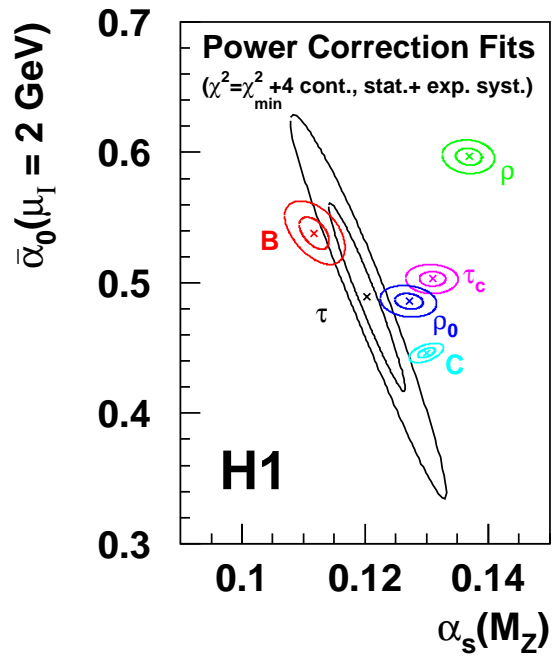, width = 0.5 \textwidth}
\caption{Summary of H1 results for fits to event shape means. Figure  taken from Ref.~\cite{Uli}.} 
\end{center}
\label{hum}
\end{figure}
\begin{figure}
\begin{center}
\epsfig{file = 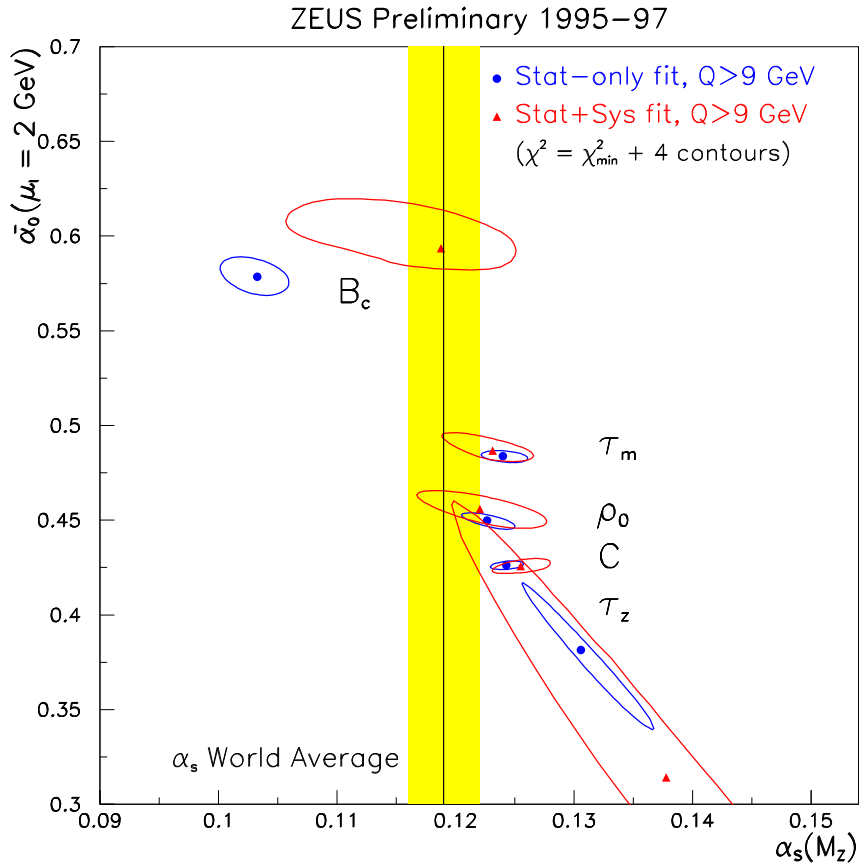, width = 0.8 \textwidth}
\caption{ZEUS contour plots for event shape means. Figure taken from \cite{Ulrike}}
\end{center}
\label{mbell}
\end{figure}
Another area of investigation at HERA is the study of event shape differential distributions $\frac{d\sigma}{dV}$ and power corrections. Here while data has been available for a while there was a lack of theoretical estimates.
Although  there are Monte Carlo programs to make fixed order estimates at $\mathcal{O}(\alpha_s^2)$ this is not sufficient in the above case.
The presence of large logarithms as big as $\alpha_s^{n} \ln^{2n-1} \frac{1}{V}$ 
at $n^{th}$ perturbative order means that a fixed order expansion is not very useful in the small $V$ region. 

As in $\ee$ one needs to resum all orders till 
next-to--leading logarithmic accuracy to describe the data.
This resummation program has been started \cite{ADS} and results for many variables are soon to be published \cite{prep}.
However a preliminary comparison to data has already begun.

The power corrections to different event shape distributions result in simple shifts of the perturbative spectra, the amount of the shift is exactly equal to the $1/Q$ correction to the corresponding mean values.
\begin{equation}
\frac{d\sigma}{dV}(V) = \frac{d\sigma^{PT}}{dV}(V+\frac{a_v\cA_1}{\pi Q})
\end{equation}
where $a_v$ for the different variables are quoted. 
\begin{center}
\begin{figure}
\epsfig{file = 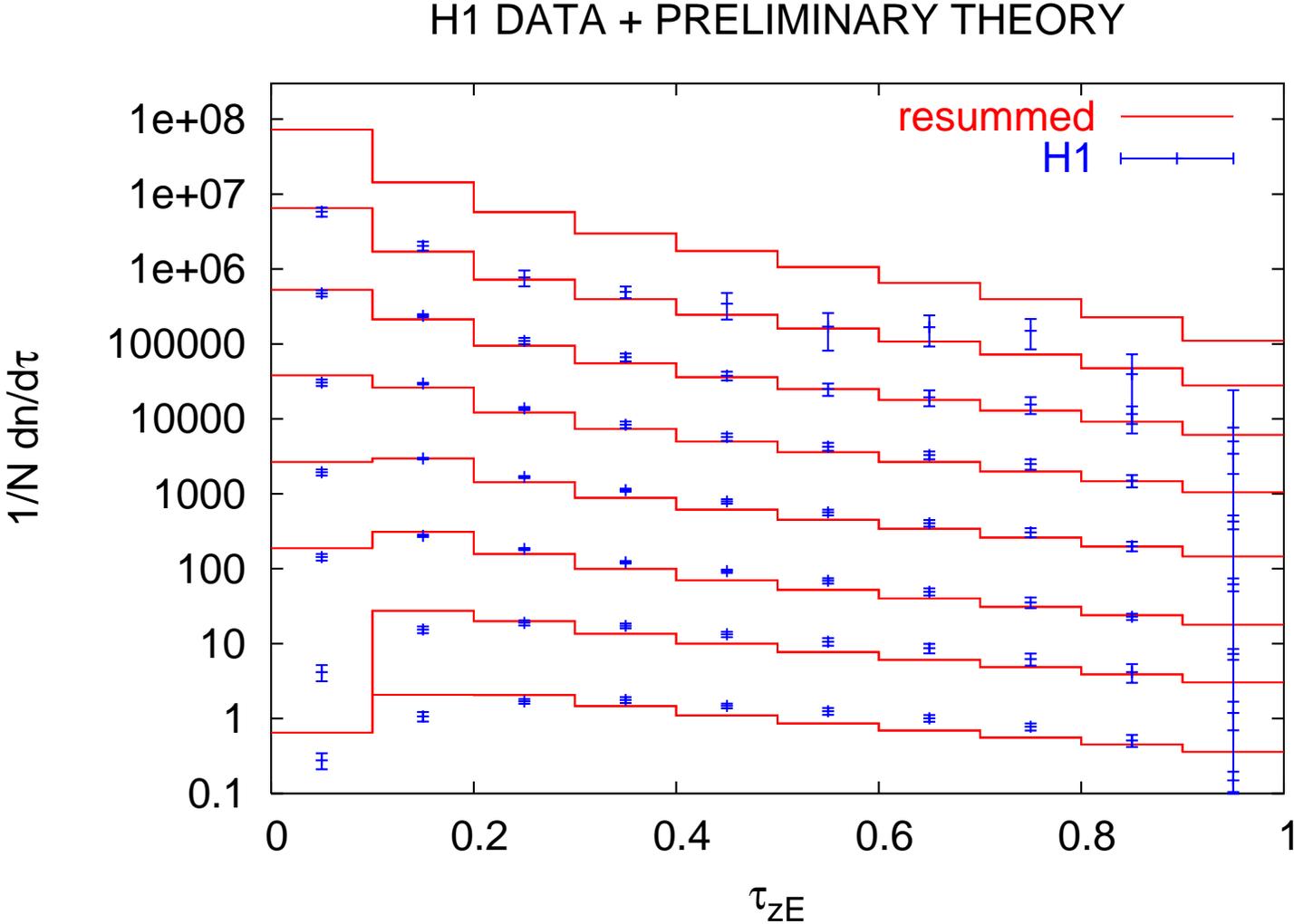, width =0.9 \textwidth}
\caption{Resummed results of \cite{ADS} supplemented by power corrections compared to H1 data. The curves from bottom to top represent increasing $Q$ values from 7 to 81 GeV.}
\end{figure}
\end{center}
Again the jet broadening is different and does not show a simple shift of the 
PT spectrum giving instead a variable dependent shift as in $e^{+}e^{-}$.
\section{Fit results}
The experimental results from H1 and ZEUS (see Figs.~1 and 2 ) have been available for quite some time for mean values whereas fits to distributions are on the way.
For the mean values the fits are best summarised by plotting error ellipses in the $\alpha_s,\bar{\alpha}_0$ plane where $\bar{\alpha_0}$ is the parameter introduced in Eq.~ \ref{A1exp}.

Some brief comments about the results are in order.
In the H1 results one should note that the jet mass ellipse is shifted and falls in line with expectations from other variables provided one neglects hadron masses (the relevant ellipse is labelled $\rho_0$) in keeping with the theoretical approximations made. For a fuller discussion of this point see \cite{gavwic}.
In the ZEUS plot one notices the effect of large systematic errors. 
ZEUS have also studied the $x$ dependence of power corrections and as noted before in this article the power correction to the jet broadening variable displays a clear $x$ dependence which comes from the anomalous dimension matrix \cite{prep}.

In addition one remarks that that the values of the $\bar{\alpha}_0$ parameter obtained from DIS event shapes has similar values (0.4--0.5) as for $e^{+}e^{-}$ variables and this seems to support the hypothesis of a universal infrared finite coupling. The spread in $\alpha_s$ values however is still uncomfortably large and is a potential area for improvement.

Finally turning to differential distributions one can use a nominal value of $\alpha_s =0.118$ and $\bar{\alpha}_0 =0.5$ and the resummation of \cite{ADS}. 
The curves in Fig.~3 are produced by binning like the H1 data which is responsible for their non-smooth nature. 
Preliminary indications are positive as shown by Fig.~3 but detailed fits are awaited.

\end{document}